\documentclass[11pt]{article}
\usepackage{moriond,epsfig}

\include{babarsym}

\def\MyPhoto{\psfig{figure=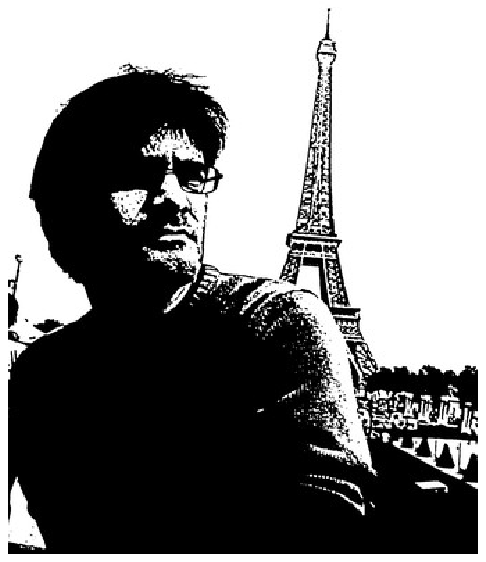,width=35mm}}

\newcommand{\lll}     {\ensuremath{\ell^{-}\ell^{+}\ell^{-}}}
\newcommand{\eee}     {\ensuremath{e^-\!e^+\!e^-}}
\newcommand{\eemw}    {\ensuremath{\mu^+\!e^-\!e^-}}
\newcommand{\eemr}    {\ensuremath{\mu^-\!e^+\!e^-}}
\newcommand{\emmw}    {\ensuremath{e^+\!\mu^-\!\mu^-}}
\newcommand{\emmr}    {\ensuremath{e^-\!\mu^+\!\mu^-}}
\newcommand{\mmm}     {\ensuremath{\mu^-\!\mu^+\!\mu^-}}
\newcommand{\Nobs}      {\ensuremath{N_{\rm obs}}}
\newcommand{\Nbgd}      {\ensuremath{N_{\rm bgd}}}
\newcommand{\Nul}       {\ensuremath{N_{\rm UL}^{90}}}
\newcommand{\BRul}       {\ensuremath{BR_{\rm UL}^{90}}}

\newcommand{\taulll}  {\ensuremath{\tau^{-}\!\to\lll}}

\newcommand{\BABARPubYear}    {08}

\newcommand{\BABARProcNumber} {199}
\newcommand{\SLACPubNumber} {13814}
\newcommand{\LANLNumber} {0000}

\def\tautau{$\tau\tau$}

\def\epem{$e^{+}e^{-}$}
\def\electron{$e$}
\def\L{\it L}

\def\be{\begin{equation}}

\def\ee{\end{equation}}
\def\bea{\begin{eqnarray}}
\def\eea{\end{eqnarray}}

\begin{document}

\begin{flushright}
SLAC-PUB-\SLACPubNumber \\
\babar-PROC-\BABARPubYear/\BABARProcNumber \\
hep-ex/\LANLNumber \\
June, 2009 \\
\end{flushright}

\vspace*{4cm}
\title{Hot Topics in \babar}

\author{ Jo$\tilde{a}$o Firmino da Costa, representing the \babar\ collaboration }

\address{ Laboratoire de l Acc$\acute{e}$l$\acute{e}$rateur Lin$\acute{e}$aire Universit$\acute{e}$ Paris-Sud 11\\
UMR 8607, B$\hat{a}$timent 200, 91898 Orsay cedex, France}

\maketitle\abstracts{
We present recent results concerning the searches for light Higgs-like particles in the decay $\Upsilon (3S) \rightarrow \gamma A^{0}, A^{0}\rightarrow \mu^{+}\mu^{-}$ as well as for the lepton flavour violation in the decays $\Upsilon (3S) \rightarrow e^{\pm}\tau^{\mp},~\mu^{\pm}\tau^{\mp}$ and $\tau \rightarrow 3l~(l=e,\mu)$  with the \babar\ experiment.
}

\section{$\Upsilon (3S) \rightarrow \gamma A^{0}, A^{0}\rightarrow \mu^{+}\mu^{-}$}
A search for a two-body transition $\Upsilon (3S) \rightarrow \gamma A^{0}$, where $A^{0}$ is a light scalar particle, followed by its decay into 2 muons has been performed using a data sample of $(121\pm1.2)\times10^{6}$ $\Upsilon (3S)$ decays. As $\Gamma_{\Upsilon(4S)}/\Gamma_{\Upsilon(3S)}\approx10^{3}$, one has greater sensitivity to rare decays using $\Upsilon(3S)$ decays than $\Upsilon(4S)$ samples.\\
The Standard Model (SM) predicts the existence of the Higgs boson to account for the different masses of elementary particles. A single SM Higgs boson is required to be heavy ($m_{H}>114.4$ GeV and $m_{H}\neq170$ GeV) \cite{ref:analysis1}. This model suffers from the quadratic divergences in the radiative corrections to the mass parameter of the Higgs potential. Other models such as the Minimal Supersymmetric Standard Model (MSSM) regulate this problem but some issues of parameter fine-tuning are not solved. A solution for the MSSM issues is to add additional Higgs fields, where one of them is naturally light. This is the case of Next-to-Minimal Supersymmetric Standard Model (NMSSM).  Direct searches constrain $m_{A^{0}}$ to be below $2m_{b}$, therefore accessible to $\Upsilon$ decays. An ideal channel\cite{ref:analysis1} is $\Upsilon \rightarrow \gamma A^{0}$, which can have a branching fraction up to $10^{-4}$ . There are already some results from CLEO \cite{ref:analysis1} which impose an upper limit for $\Upsilon(1S) \rightarrow \gamma A^{0}(\mu^{+}\mu^{-})\approx 4\times 10^{-6}$ for $m_{A^{0}}<2m(\tau)$.\\
 Other physical models, motivated by astrophysical observations, predict similar light states. For instance, \cite{ref:analysis1} proposes a light axion-like particle $a$ decaying predominantly to leptons and predicts a branching fraction $\Upsilon (3S) \rightarrow \gamma a$ to be between $10^{-6}-10^{-5}$. Another motivation for a light Higgs like particle comes from the HyperCP experiment \cite{ref:analysis1} which has observed three anomalous events for the decay $\Sigma \rightarrow p \mu^{+}\mu^{-}$. These have been interpreted as a light scalar with mass of $214.3$ MeV decaying into a pair of muons.  Finally we look for leptonic decays of the $\eta_{b}$ in $\Upsilon(3S)$ decays for better understanding of this newly found state. One expects its leptonic width to be negligible if this is a conventional quark-antiquark state.

\begin{figure}[t]
	\centering
		\includegraphics[width=8cm,height=6cm]{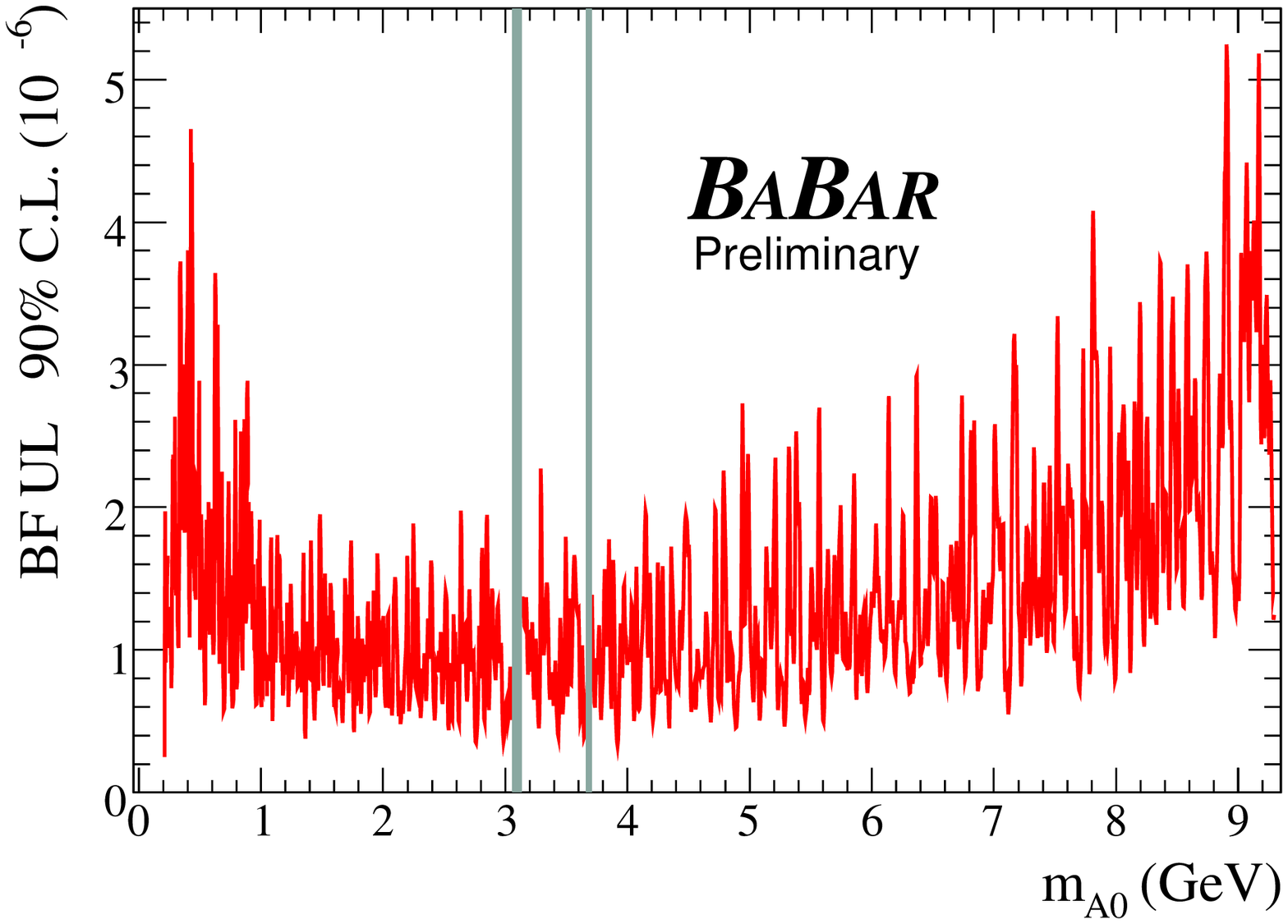}\includegraphics[width=8cm,height=6cm	]{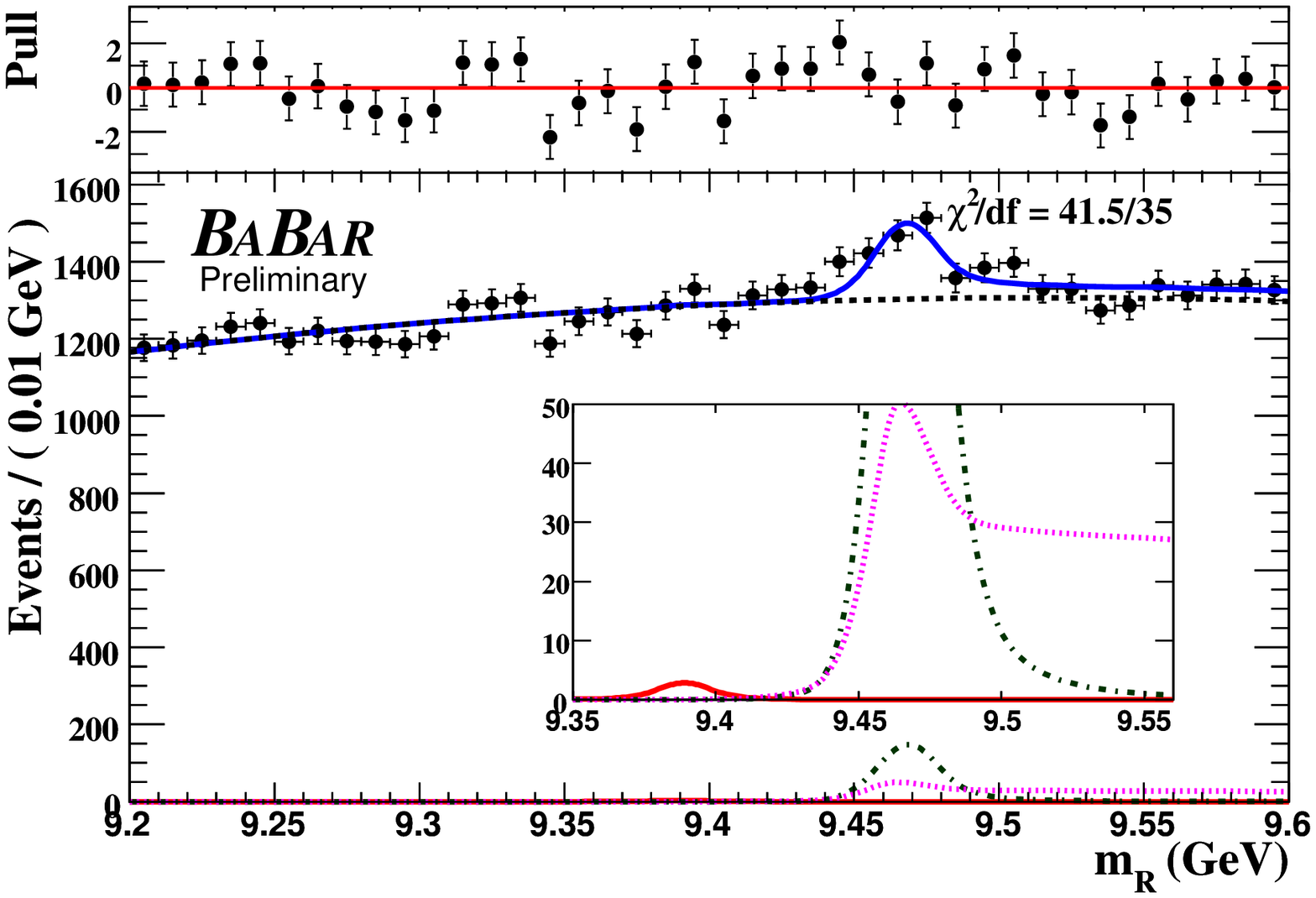}
	\label{fig:CL_A0}
\caption{Left plot: Upper limits on the product of branching fractions $B(\Upsilon(3S)\rightarrow \gamma A^{0})\times B(A^{0}\rightarrow \mu^{+}\mu^{-})$ as a function of $m_{A^{0}}$ from the
fits to $\Upsilon(3S)$ data. The shaded areas show the regions around the $J/\psi$  and  $\psi(2S)$ resonances excluded from the search. Right plot: The fit for the $\eta_{b}$ region in $\Upsilon(3S)$ dataset. The bottom graph shows the mR distribution (solid points), overlaid by the full PDF (solid blue line). Also shown are the contributions from the signal at $m_{\eta_{b}} = 9.389$ GeV (solid red line), background from the \epem $~\rightarrow \gamma_{ISR}\Upsilon(1S)$ (dot-dashed green line), background from $\Upsilon(3S)\rightarrow \chi_{b}(2P)$,$\chi_{b}(2P)\rightarrow  \gamma \Upsilon(1S)$
(dotted magenta line), and the continuum background (dashed black line). The inset shows the signal, 
\epem $\rightarrow \gamma_{ISR}\Upsilon(1S)$ and $\chi_{b}(2P)\rightarrow  \gamma \Upsilon(1S)$ in more detail. The top plot shows the normalized residuals p = $(data - fit)/\sigma(data)$ with unit error bars.}
\end{figure}

%Strategy and method
The event selection is explained elsewhere \cite{ref:analysis1}. We extract the yield of signal events as a function of the assumed mass $m_{A^{0}}$ in the interval $0.212\leq m_{A^{0}}<9.3$ GeV by performing a series of unbinned extended maximum likelihood fits to the distribution of the "reduced mass" 
\begin{equation}
m_{R}=\sqrt{m_{\mu\mu}^{2}-4m_{\mu}^{2}}.
\end{equation}
The advantage of this variable is that the main background is  smooth throughout the entire range of interest. Each fit is performed over a small range of $m_{R}$ around a particular value of $m_{A^{0}}$. The signal PDF for each fit is determined from simulated signal events generated at many different mass points. For fits done between two of these mass points, the PDF shape is interpolated. Simulation results are calibrated using charmonium backgrounds. The background PDF is determined from fit to $\Upsilon(4S)$ data. The background is dominated by \epem $\rightarrow \gamma \mu^{+} \mu^{-}$ either from continuum or through ISR-produced $J/\psi,\psi(2S),\Upsilon(1S)$.\\
%Results
We scan almost 2000 mass points and for each we determine the significance of any particular peak. The probability to find a 3 $\sigma$ effect due to statistical fluctuation is of about 80$\%$. The most significant peak is at $m_{A^{0}}=4.940\pm0.003~GeV$ (=$3\sigma$). The second most-significant peak is at $m_{A^{0}}=0.426\pm0.001~GeV$(=$2.9\sigma$).  . As a consequence of not finding significant excess of events above the background in the range $0.212<m_{A^{0}}<9.3$ GeV, we set upper limits on the branching fraction $B(\Upsilon(3S) \rightarrow \gamma A^{0})\times B(A^{0}\rightarrow \mu^{+}\mu^{-})$. The 90$\%$ C.L. Bayesian upper limits, computed with a uniform prior and assuming a Gaussian likelihood are shown in Figure \ref{fig:CL_A0} as a function of the mass $m_{A^{0}}$. The limits fluctuate depending on the central value of the signal yield returned by a particular fit, and range from $0.25\times10^{-6}$ to $5.2\times10^{-6}$.\\

No significant signal is observed at the HyperCP mass (m=0.214 GeV). We set an upper limit $B(\Upsilon(3S) \rightarrow \gamma A^{0}(214))<0.8 \times 10^{-6}$ at $90\%$ C.L. No significant signal is observed in the $\eta_{b}$ region (Figure  \ref{fig:CL_A0}) and combining these results with the \babar $~$ measurement of $B(\Upsilon(3S)\rightarrow \gamma \eta_{b}) = (4.8 \pm 0.5 \pm1.2)\times 10^{-4}$ we can impose an upper limit to $B(\eta_{b}\rightarrow \mu^{+}\mu^{-})<0.8\%$ at 90$\%$ C.L. This is consistent with a quark-antiquark view of this state. All results presented above are preliminary.
The limits we set are more stringent than those reported by the CLEO collaboration. Our limits rule
out much of the parameter space allowed by the light Higgs and axion models. Namely, most of the parameter space is excluded for $m_{A^{0}}<2m_{\tau}$. For masses between this value and the $\Upsilon(3S)$ it is more relevant to observe $\tau$ and hadronic decay modes. These searches are ongoing.

\section{$\Upsilon(3S) \rightarrow e^{\pm}\tau^{\mp},\mu^{\pm} \tau^{\mp}$}
\begin{figure}[!b]
	\centering
		\includegraphics[width=14cm]{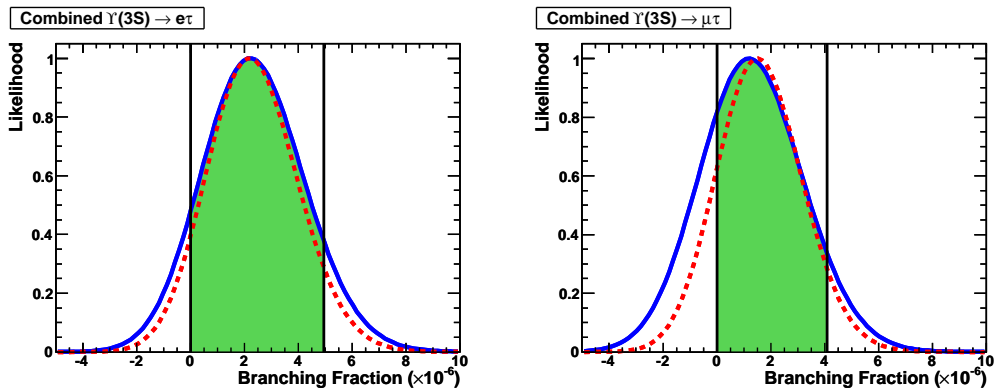}
	\label{fig:upsilontoetaumutau}
\caption{The results of the likelihood scan. The likelihood curves for the two  $\Upsilon(3S)\rightarrow e^{\pm}\tau^{\mp}$ and two $\Upsilon(3S)\rightarrow \mu^{\pm}\tau^{\mp}$ signal channels have been multiplied to obtain the combined likelihood curves. The red dashed lines indicate statistical uncertainties only, the solid blue lines have systematic uncertainties incorporated. The vertical lines bound $90\%$  of the positive integral of the total likelihood curve, which is shaded in green.}
\end{figure}

A search for charged lepton-flavor violation (CLFV) in the $\Upsilon$ sector is presented. We use a sample of $(116.7\pm1.2)\times 10^{6}$ $\Upsilon(3S)$ decays corresponding to a luminosity of 27.5 $fb^{-1}$.\\
The search for CLFV has been intense in the $\mu$ and $\tau$ sector but the $\Upsilon$ sector is less analyzed. If new particles contributing to CLFV is in the Higgs area, it would rather couple to heavy quark flavours, and if the new physics couples to b quarks, this may be observable in decays of the $\Upsilon$. There are prior constraints to $\Upsilon$ CLFV from CLEO \cite{ref:analysis2} which places a $95\%$ C.L. upper limit for $B(\Upsilon(3S) \rightarrow \mu^{\pm} \tau^{\mp})<20.3\times 10^{-6}$. 
 The event selection is explained in detail elsewhere \cite{ref:analysis2}. The $\tau$ is reconstructed leptonically (its daughter lepton must not have the same flavour as the other lepton of the $\Upsilon$ decay) or hadronically ( we require one or more $\pi^{0}$'s from this decay). The main source of background comes from $\tau$-pair production, which is dominated by continuum production, although a non negligible contribution from $\Upsilon(3S)\rightarrow \tau^{+}\tau^{-}$, Bhabha and $\mu$-pair production exists.\\
The discriminant variable is the beam-energy-normalized primary (electron or muon) CM momentum $x=p_{1}/E_{B}$. The fit is done using an unbinned extended maximum likelihood where we define 3 PDFs (signal, $\tau$-pair production, BhaBha, $\mu$-pair production), where the yields for each component are free parameters. These PDFs are determined from simulation and calibration data.\\
The fit results are all consistent with zero within $2.1~\sigma$. The  branching fractions is scanned in small increments. Each fit is performed with the signal yield (and consequently the branching fraction) fixed at a given value. We present this likelihood scan in Figure \ref{fig:upsilontoetaumutau}.\\
The upper limits are determined to be: $BF(\Upsilon(3S) \rightarrow e \tau) < 5.0\times 10^{-6}$, which is the first upper limit determined for this channel and $BF(\Upsilon(3S) \rightarrow \mu \tau) < 4.1\times 10^{-6}$, which is over 4 times lower than the previous upper limit. Both of these limits are defined with $90\%$ C.L. \\   
We can impose lower limits on the mass scales of beyond-Standard model physics that contribute to LFV in $\Upsilon(3S)$  decays. This is done in the framework of an effective field theory where the new physics effects are characterized by a $\Lambda_{N}$ parameter (mass scale) and a coupling constant $\alpha_{N}$. Assuming a strong coupling ($\alpha_{N}$=1) our branching fractions upper limits correspond to lower limits in the mass scale  $\Lambda_{N}^{e\tau}>1.4$ TeV and $\Lambda_{N}^{e\tau}>1.5$ TeV respectively at 90$\%$ C.L.	These are more stringent than already existing limits \cite{ref:analysis2}. 

\section{$\tau \rightarrow 3l~(l=e,\mu)$}
A search for the neutrinoless, lepton-flavor violating decay of the 
tau lepton into three charged leptons has been performed  using 472 $fb^{-1}$ of BaBar data.
This search is an update of a previous BaBar analysis\cite{ref:analysis3}, using a 
different selection technique, and an improved particle identification.\\

Lepton-flavor violation (LFV) involving charged leptons has 
never been observed, and stringent experimental limits 
exist from muon branching fractions\cite{ref:analysis3}:
$BR(\mu\to\electron\gamma) < 1.2 \times 10^{-11}$ 
and $BR(\mu\rightarrow\electron\electron\electron) < 1.0 
\times 10^{-12}$  at 90\% CL.
\taulll{} in SM are predicted to be at most similar to the $\tau\rightarrow\mu\gamma$, which are well below experimental sensitivities.
Many descriptions of physics beyond the Standard Model (SM), 
particularly models 
seeking to describe neutrino mixing, predict enhanced LFV in $\tau$ 
decays over $\mu$ decays with branching fractions from 
$10^{-10}$ up to the current experimental limits 
\cite{ref:analysis3}.
An observation of LFV in $\tau$ decays would be a 
clear signature of non-SM physics, while improved 
limits will provide further constraints on theoretical models. 
We consider all possible lepton combination consistent with charge
conservation.\\

The event is divided into two hemispheres 
in the \epem\ center-of-mass (c.m.) frame 
using the plane perpendicular to the thrust axis,
as calculated from the observed tracks and neutral energy deposits.
The signal hemisphere must contain exactly three tracks (each identified as either an electron or muon,
with an invariant mass and energy equal to that of the parent 
tau lepton), while the other hemisphere must contain exactly one. The tracks are required to point toward a common region consistent with \tautau\ production and decay.
We also require that the three tracks from the signal hemisphere should
have an invariant mass of less than 3.5$~$\gevcc.
At this preselection stage all charged particles are assumed to be pions, and 
so the preselection efficiency is lower for channels with electrons.\\

Particle identification (PID) uses new algorithms relative to those used
in \cite{ref:analysis3} which greatly 
improve electron, and in particular muon identification capabilities with
respect to the previous analysis.
These improvements result in an average efficiency per lepton of 91\% and 77\% respectively 
(older PID had a 65\% efficiency for muons).
The probability for a pion to be misidentified as an electron in 
3-prong tau decays is 2.4\%, while the probability to be misidentified
as a muon is 2.1\%. Older selectors have misidentification rates
of 2.7\% for electrons, and 2.9\% for muons respectively.\\

For further background reduction,
candidate signal events are required to have
an invariant mass and total energy in the 3-prong
hemisphere consistent with a parent tau lepton.
These quantities are calculated from the observed track momenta 
assuming lepton masses that correspond to the specific decay mode.
The quantity
$\Delta E \equiv E^{\star}_{\mathrm{rec}} - E^{\star}_{\mathrm{beam}}$,
is defined, where $E^{\star}_{\mathrm{rec}}$ is the total energy of the tracks 
observed in the 3-prong hemisphere and $E^{\star}_{\mathrm{beam}}$
is the beam energy, with both quantities measured in the c.m. frame.
The quantity
$\Delta M_{rec} \equiv M_{\mathrm{rec}} - m_{\tau}$ is defined with 
$M_{\mathrm{rec}}^{2}\equiv E^{\star \, 2}_{beam}/c^{4} - |\vec{p}_{3l}^{\, \star}|^{2}/c^{2}$, 
where $|\vec{p}_{3l}^{\, \star}|^{2}$ 
is the squared momentum of the 3-prong tracks in the c.m. and 
$m_{\tau}=1.777\gevcc$ is the tau mass.

\begin{table}
\begin{center}
\caption{Efficiency estimates, number of expected background events (\Nbgd),
expected branching fraction upper limits at 90\% CL (UL$_{90}^{\rm exp}$),
number of observed events (\Nobs), and observed branching fraction upper limits 
at 90\% CL (UL$_{90}^{\rm obs}$) for each decay mode. All upper limits are in
units of $10^{-8}$.
}
\begin{tabular}{lccccc}
\hline\hline
Mode & Eff. [\%] & \Nbgd  & UL$_{90}^{\rm exp}$ & \Nobs & \rule{0pt}{13pt}UL$_{90}^{\rm obs}$ \\
\hline
\eee  &$ 8.6  \pm 0.2 $&$ 0.12 \pm 0.02 $&$ 3.4  $&$ 0 $&$ 2.9 $\\ 
\eemr &$ 8.8  \pm 0.5 $&$ 0.64 \pm 0.19 $&$ 3.7  $&$ 0 $&$ 2.2 $\\
\eemw &$ 12.7 \pm 0.7 $&$ 0.34 \pm 0.12 $&$ 2.2  $&$ 0 $&$ 1.8 $\\
\emmw &$ 10.2 \pm 0.6 $&$ 0.03 \pm 0.02 $&$ 2.8  $&$ 0 $&$ 2.6 $\\	
\emmr &$ 6.4  \pm 0.4 $&$ 0.54 \pm 0.14 $&$ 4.6  $&$ 0 $&$ 3.2 $\\
\mmm  &$ 6.6  \pm 0.6 $&$ 0.44 \pm 0.17 $&$ 4.0  $&$ 0 $&$ 3.3 $\\
\hline
\hline
\end{tabular}
\label{tab:results_tau}
\end{center}
\end{table}

The numbers of events observed (\Nobs) and the 
background expectations (\Nbgd) are shown in Table~\ref{tab:results_tau}, 
with no events found in any decay mode.
Upper limits 
on the branching fractions are calculated according to 
$\BRul = \Nul/(2 \varepsilon \textsl{L} \sigma_{\tau\tau})$, where $\Nul$
is the 90\% CL upper limit for the number 
of signal events when \Nobs\ events are observed with \Nbgd\ background 
events expected.
The values $\varepsilon$, $\L$, and $\sigma_{\tau\tau}$ are the
selection efficiency, luminosity, and \tautau\  cross section, respectively.
The uncertainty on the product $\L \cdot \sigma_{\tau\tau}$ is 0.9\%. 
The sensitivity or expected upper limit UL$_{90}^{\rm exp}$, 
defined as the mean upper limit expected in the background-only hypothesis, 
is included in Table~\ref{tab:results_tau}.
The 90\% CL upper limits on the \taulll\ branching fractions 
are in the range $(1.8-3.3)\times10^{-8}$.
This analysis supersedes the previous \babar\ analysis \cite{ref:analysis3}.
These results represent also an improvement with respect to the previous
experimental bounds \cite{ref:analysis3}, 
obtaining smaller UL for \eee, \eemr, \eemw, and \emmr channels.
This analysis has improved the UL by a factor $\approx5-8$ with only a small increase in luminosity thanks to the use of new tools and new event selections.

\section*{Acknowledgments}
I would like to thanks Alberto Cervelli, Alberto Lusiani,Benjamin Hooberman and Yury Kolomensky for their important feedback for this presentation and their availability. I thank the Moriond EW organizing committee for the excellent organisation.

\end{document}